\documentclass[11pt,reqno]{amsart}      

\usepackage{amssymb,amsthm}             

\usepackage{eucal}                      

\usepackage{calc}			
\usepackage{graphicx}
\graphicspath{pic/}			
\usepackage{epic,eepic}			

\newcommand{\mnote}[1]{} 

\renewcommand{\Re}{{\mathbb R}}         
\newcommand{\la}{\langle}               
\newcommand{\ra}{\rangle}               
\newcommand{\half}{\frac{1}{2}}         
\newcommand{\tr}{\text{\rm tr}}		
\newcommand{\Area}{\mathcal{A}}		
\newcommand{\Energy}{\mathcal{E}}	
\newcommand{\norm}{\eta} 		
\newcommand{\Hor}{\mathcal H}		
 		
\newcommand{\Ric}{\text{\rm Ric}}       
\newcommand{\Scal}{\text{\rm Scal}}     
\theoremstyle{plain}
\newtheorem{thm}{Theorem}

\title{Geometric Analysis and General Relativity}
\author[L. Andersson]{Lars Andersson$^1$}
\thanks{$^1$Supported in part by the NSF under
contract no. DMS 0407732 with the University of Miami.
}
\address{Department of Mathematics\\
University of Miami\\
Coral Gables, FL 33124\\
USA \and
Albert Einstein Institute\\
Am M\"uhlenberg 1
D-14476 Golm
Germany}
\email{larsa\char'100math.miami.edu}
\date{August 31, 2005}

\begin{document}
\maketitle

Geometric analysis can be said to originate in the 19'th century 
work of Weierstrass, Riemann, Schwarz and 
others on minimal surfaces, a problem whose history can be traced at least as
far back as the work of Meusnier and Lagrange in the 18'th century. 
The experiments performed by Plateau in the mid-19'th century, 
on soap films spanning wire contours, 
served as an important
inspiration for this work, and let to the formulation of the Plateau
problem, which concerns the existence and regularity of area minimizing
surfaces in $\Re^3$ spanning a given boundary contour. The Plateau problem
for area minimizing disks spanning a curve in $\Re^3$ was solved by 
Jesse Douglas (who shared the first Fields medal with Lars V. Ahlfors) 
and Tibor Rado in the 1930's. 
Generalizations of Plateau's problem 
have  been an important driving force behind the
development of modern geometric analysis. Geometric analysis 
can be viewed broadly as the
study of partial differential equations arising in geometry, and includes 
many areas of the calculus of variations, 
as well as 
the theory of geometric evolution equations. The Einstein equation, which
is the central object of general relativity, is one of the most widely studied 
geometric partial differential equations, and plays an important role in
its Riemannian as well as in its Lorentzian form, the Lorentzian being most
relevant for general relativity.

The Einstein equation is the Euler-Lagrange equation of a
Lagrangian with gauge symmetry and thus in the Lorentzian case it, 
like the Yang-Mills equation, can be viewed as 
a system of evolution equations with constraints. After imposing
suitable gauge conditions, the Einstein equation becomes a hyperbolic
system, in particular using spacetime harmonic coordinates (also known as
wave coordinates), the Einstein equation becomes a quasilinear system of wave
equations. The constraint equations implied by the Einstein equations 
can be viewed as a system of elliptic
equations in terms of suitably chosen variables. 
Thus the Einstein equation 
leads to both elliptic and hyperbolic problems, arising from the
constraint equations and the Cauchy problem, respectively. The groundwork for
the mathematical 
study of the Einstein equation and the global nature of spacetimes was
laid by, among others, Choquet-Bruhat who proved local 
well-posedness for the Cauchy
problem, 
\mnote{role of ChB in constraint problem?} 
Lichnerowicz, and later York who provided the basic ideas for the
analysis of the constraint equations, and Leray who formalized the notion of
global hyperbolicity, which is essential for the global study of spacetimes. 
An important framework for the mathematical study of the Einstein
equations has been provided by the singularity theorems of Penrose and
Hawking, as well as the cosmic censorship conjectures of Penrose. 

Techniques and ideas from
geometric analysis have played, and continue to play,
a central role in recent mathematical progress
on the problems posed by general relativity. 
Among the main results are the proof of the positive mass theorem using the
minimal surface technique of Schoen and Yau, and the spinor based approach of
Witten, as well as 
the proofs of the
(Riemannian) Penrose inequality by Huisken and Illmanen, and Bray. 
The proof of the Yamabe theorem by Schoen has played an important role as a
basis for constructing Cauchy data using the conformal method. 

The results just mentioned are all essentially 
Riemannian in nature,
and do not involve study of the Cauchy problem for the Einstein equations. 
There has been great progress recently concerning 
global results on the Cauchy problem for the Einstein equations, 
and the cosmic censorship conjectures of Penrose. The results available so
far are either small data results, among these the nonlinear 
stability of Minkowski space proved by Christodoulou and Klainerman, or
assume additional symmetries, such as the recent proof by Ringstr\"om of 
strong cosmic censorship for the class of Gowdy spacetimes. 
However, recent progress concerning
quasilinear wave equations and the geometry of spacetimes with low regularity
due to, among others, Klainerman and Rodnianski, and Tataru and Smith, appears
to show the way towards an improved understanding of the Cauchy problem for
the Einstein equations. 

Since the constraint equations, the Penrose inequality and the 
Cauchy problem are discussed
in separate articles, the focus of this article will be on 
the role in general relativity 
of ``critical'' and other geometrically defined submanifolds and
foliations,  such as minimal surfaces, marginally
trapped surfaces, constant mean curvature hypersurfaces and null
hypersurfaces. 
In this context it would be natural also 
to discuss geometrically defined flows such as
mean curvature flows, inverse mean curvature flow, and Ricci flow. However,
this article restricts 
the discussion to mean curvature flows, since the inverse
mean curvature flow appears naturally in the context of the Penrose
inequality 
and the Ricci flow has 
so far mainly served as a source of inspiration for research on the Einstein
equations rather than an important tool. 
Other topics which would fit well under the heading ``General
Relativity and Geometric
Analysis'' are spin geometry (the Witten proof of the Positive mass theorem),
the Yamabe theorem and related results concerning the Einstein constraint
equations, gluing and other techniques of ``spacetime engineering''. These
are all discussed in other articles. Some
techniques which have only recently come into use and for which 
applications in general relativity have not been much explored, 
such as Cheeger-Gromov compactness, are not discussed.

\section{Minimal and related surfaces} \label{sec:minimal} 
Consider a hypersurface $N$ in 
Euclidean space $\Re^{n}$ 
which is a graph $x_n = u(x_1,\dots,x_{n-1})$ with
respect to the function $u$. 
The area of $N$ is given by 
$\Area(N) = \int \sqrt{1 + |D u|^2} dx^1\cdots dx^{n-1}$. 
$N$ is stationary with respect to $\Area$ 
if $u$ satisfies the equation 
\begin{equation}\label{eq:graph}
\sum_i D_i \left (\frac{D_i u}{\sqrt{1 + |D u|^2}} \right ) = 0
\end{equation} 
A hypersurface $N$ defined as a graph of $u$ solving (\ref{eq:graph})
minimizes area with respect to compactly supported deformations, and hence is
called a {\em minimal surface}. 
For $n \leq 7$, a solution to equation (\ref{eq:graph}) 
defined on all of $\Re^{n-1}$ must be an affine function. 
This fact is known as a
Bernstein principle. 
Equation (\ref{eq:graph}), and more generally, the
prescribed mean curvature equation which will be discussed below, 
is a quasilinear, uniformly
elliptic second order equation. The book \cite{GT} is an excellent general reference for
such equations. 

The theory of rectifiable currents, developed by Federer and Fleming, is a
basic tool in the modern approach to the Plateau problem and related
variational problems. A rectifiable current 
is a countable union of Lipschitz submanifolds, counted with integer 
multiplicity, and satisfying certain regularity conditions.
Haussdorff measure gives a  notion of area for these objects. One may
therefore approach the study of minimal surfaces via 
rectifiable currents which are stationary with respect to variations of
area. Suitable
generalizations of familiar notions from smooth differential geometriy such
as tangent plane, normal vector, extrinsic
curvature can be introduced.  
The book by Federer \cite{federer:book} is a
classic treatise on the subject. 
Further information concerning minimal surfaces
and related variational problems can be found in 
\cite{lawson:minsurf,simon:survey}. 
Note, however, that unless otherwise
stated, all fields and manifolds considered in this article 
are assumed to be smooth. 
For the Plateau problem in a Riemannian ambient space, we have the following
existence and regularity result.
\begin{thm}[Existence of embedded solutions for Plateau problem]
Let $M$ be a complete Riemannian manifold of dimension $n \leq 7$ and let 
$\Gamma$ be a compact $n-2$ dimensional submanifold in $M$ which bounds. 
Then there is an $n-1$ dimensional area minimizing 
hypersurface $N$ with $\Gamma$ as its boundary. $N$ is a smooth,
embedded manifold in its interior. 
\end{thm} 
If the dimension of the ambient space is greater than 7, 
solutions to the Plateau problem will in general have a singular set of
dimension $n-8$. 
Let $N$ be an oriented hypersurface of a Riemannian 
manifold $M$ 
with covariant derivative $D$. 
Let $\norm$ be the unit normal of $N$ and define the second fundamental
form and mean curvature of $N$ by $A_{ij} = \la D_{e_i} \norm, e_j \ra$ and $H
= \tr A$. 
Define the action functional $\Energy(N) = \Area(N) 
- \int_{M;N} H_0$, where $H_0$ is a function defined on $M$, and
$\int_{M;N}$ denotes the integral over the volume bounded by $N$ in
$M$. 
The problem of minimizing $\Energy$ is a useful
generalization of the minimization problem for $\Area$. 
\begin{thm}[Existence of minimizers in homology] 
\mnote{cite Federer [federer:book]} 
Let $M$ be a compact Riemannian manifold of dimension $\leq 7$, and let
$\alpha$ be an integral homology class on $M$ of codimension one. Then there
is a smooth minimizer for 
$\Energy$ 
representing $[\alpha]$. 
\mnote{also if $M$ has dimension 3 and $N$ is a surface not the sphere, then 
given $f: N \to M$ incompressible, there is a minimizer freely homotopic to
$f(N)$, see [Auer]}
\end{thm} 
Again, in higher dimensions, the minimizers will in general have
singularities. The general form of this result deals with elliptic
functionals. For surfaces in 3-manifolds, the problem of minimizing area
within homotopy classes has been studied. Results in this direction played a
central role in the approach of Schoen and Yau to manifolds with non-negative
scalar curvature. \mnote{see Sacks-Uhlenbeck, [Auer] for comments on this} 

If $M$ is not compact, it is in general necessary to use barriers to control
the minimizers, or consider some version of the Plateau problem. 
Barriers can
be used due to the strong maximum principle, which holds for the 
mean curvature operator since it is quasilinear elliptic.
Consider two 
hypersurfaces $N_1, N_2$ which intersect in a point $p$ and assume that $N_1$
lies on one side of $N_2$ with the normal pointing towards $N_1$. 
If the mean
curvatures $H_1, H_2$ of the hypersurfaces, 
defined with respect to consistently oriented
normals, satisfy $H_1 \leq \lambda \leq H_2$ for some constant $\lambda$, 
then $N_1$ and $N_2$ coincide
near $p$ and have mean curvatures equal to $\lambda$. This result
requires
only mild regularity conditions on the hypersurfaces. Generalizations hold 
also for the case of spacelike or null hypersurfaces in a Lorentzian ambient
space,  
see \cite{andersson:etal:maxprinc,galloway:nullmax}.

Let $\phi$ be a smooth compactly supported function on $N$.
The variation $\Energy' = \delta_{\phi \norm} \Energy$ of 
$\Energy$ under a deformation $\phi \norm$ is 
$$
\Energy '  =  \int_N \phi (H - H_0)
$$
Thus $N$ is stationary with respect to $\Energy$ if and only if $N$ solves the
prescribed mean curvature equation $H(x) = H_0(x)$
for $x \in N$. 
Supposing that $N$ is stationary and $H_0$ is constant, 
the second variation $\Energy'' = \delta_{\phi \norm} \Energy'$ 
of $\Energy$ is of the form 
$$
\Energy ''  = \int_N \phi (J \phi)
$$
where $J$ is the second variation operator, a second order elliptic operator.
A calculation, using the Gauss equation and the second variation equation shows 
\begin{equation}\label{eq:J} 
J \phi =  -  \Delta_N \phi 
- \half [  (\Scal_M - \Scal_N) + H^2 +
|A|^2 ] \phi ,
\end{equation}
where $\Delta_N, \Scal_M, \Scal_N$ denote the Laplace-Beltrami operator of
$N$, and the scalar curvatures of $M$ and $N$, respectively. 
If $J$ is positive semidefinite, 
$N$ is called stable.



To set the context where we will apply the above, let $(M,g_{ij})$ be a
connected, asymptotically Euclidean three-dimensional Riemannian manifold
with covariant derivative, and let $K_{ij}$ be a symmetric tensor on $M$. 
Suppose $(M,g_{ij},
K_{ij})$ is imbedded isometrically as a spacelike hypersurface in a spacetime
$(V, \gamma_{\alpha\beta})$ with $g_{ij}, K_{ij}$ the first and second
fundamental forms induced on $M$ from $V$, in particular 
$K_{ij} = \la D_{e_i} T, e_j \ra$ 
where $T$ is the timelike normal
of $M$ in the ambient spacetime $V$, 
and $D$ is the ambient covariant derivative. We will refer to $(M,g_{ij},
K_{ij})$ as a Cauchy data set for the Einstein equations. Although many of
the results which will be discussed below generalize to the case of a nonzero
cosmological constant $\Lambda$, 
we will discuss only the case $\Lambda = 0$ in this article. 
$G_{\alpha\beta} = {\Ric_V}_{\alpha\beta} - \half \Scal_V \gamma_{\alpha\beta}$ 
be the Einstein tensor of
$V$, 
and let $\rho = G_{\alpha\beta} T^\alpha T^\beta$, $\mu_j =
G_{j\alpha} T^\alpha$. Then the fields $(g_{ij}, K_{ij})$ satisfy the  
Einstein constraint equations
\begin{align}
R + \tr K^2 - |K|^2 &= 2\rho\\
\nabla_j \tr K - \nabla^i K_{ij} = \mu_j 
\end{align} 

We assume that the dominant energy condition (DEC)
\begin{equation}\label{eq:DEC}
\rho \geq ( \sum_i \mu_i \mu^i
)^{1/2}
\end{equation}
holds. We will sometimes make use of the null energy condition (NEC),
$G_{\alpha\beta}L^\alpha L^{\beta} 
\geq 0$ for null vectors $L$, and the strong energy condition (SEC),
$\Ric_{V\, \alpha\beta} v^\alpha v^\beta \geq 0$ for causal vectors $v$. 
$M$ will be assumed to satisfy the fall-off conditions
\begin{subequations}\label{eq:asymptcond}
\begin{align} 
g_{ij} &= (1 + \frac{2m}{r} )\delta_{ij} + O(1/r^2)  \label{eq:asymptg} \\
K_{ij} &= O(1/r^2) 
\end{align}
\end{subequations} 
as well as suitable conditions for the fall-off of derivatives of $g_{ij},
K_{ij}$. 
Here $m$ is the ADM (Arnowitt, Deser, Misner) mass of $(M,g_{ij},K_{ij})$.

\subsection{Minimal surfaces and positive mass} 
Perhaps 
the most important application of the theory of minimal surfaces in general
relativity is in the Schoen-Yau proof of the positive mass theorem, which
states that $m \geq 0$, and $m=0$ only if $(M,g,K)$ can be embedded as a
hypersurface in Minkowski space. 
Consider an asymptotically Euclidean manifold
$(M,g)$ with $g$ satisfying (\ref{eq:asymptg}) and with 
non-negative scalar curvature. 
By using Jang's equation, see below, the general situation
is reduced to the case of a time symmetric data set, with $K=0$. In this case
the DEC implies that $(M, g)$ has nonnegative scalar curvature. 

Assuming $m < 0$ one may, after applying a 
conformal deformation, assume that $\Scal_M > 0$ in the complement of a
compact set. Due to the asymptotic conditions, level sets for sufficiently
large values of one 
of the coordinate
functions, say $x^3$, can be used as barriers for minimal surfaces in $M$. 
By solving a sequence of Plateau problems with
boundaries tending to infinity, a stable entire minimal surface $N$
homeomorphic to the plane
is constructed. Stability implies using (\ref{eq:J}),
$$
\int_N (\half \Scal_M - \kappa + \half |A|^2 ) \leq 0 ,
$$
where $\kappa = \half \Scal_N$ is the Gauss curvature of $N$. 
Since by construction 
$\Scal_M \geq 0$, $\Scal_M > 0$ outside a compact set, this gives 
$\int_N \kappa > 0$. 
Next, one uses the identity, related to the Cohn-Vossen inequality 
$$
\int_N \kappa = 2\pi - \lim_i \frac{L_i^2}{2A_i}
$$
where $A_i, L_i$ are the area and circumference of a sequence of large
discs. Estimates using the fact that $M$ is asymptotically Euclidean 
show that $\lim_i \frac{L_i^2}{2A_i} \geq 2\pi$ which gives a contradiction
and shows that the minimal surface constructed cannot exist. It follows that
$m \geq 0$. It remains to show that the case $m=0$ is rigid. To do this proves
that for an asymptotically Euclidean metric with non-negative scalar
curvature, which is positive near infinity, there is a
conformally related metric with vanishing scalar curvature and strictly
smaller mass. Applying this argument in case $m=0$ gives a contradiction to
the fact that $m \geq 0$. Therefore $m=0$ only if the scalar curvature
vanishes identically. Suppose now that $(M, g)$ has vanishing scalar
curvature but non-vanishing Ricci
curvature $\Ric_M$. Then using a deformation of $g$ in the direction of
$\Ric_M$, one constructs a metric close to $g$ with negative mass, which
leads to a contradiction. 

This technique generalizes to Cauchy surfaces of dimension $n \leq
7$. 
The proof involves induction on dimension. For
$n > 7$ minimal hypersurfaces are singular in general and this
approach runs into problems. The Witten proof using spinor techniques does
not suffer from this limitation but instead requires that $M$ be spin. 
\subsection{Marginally trapped surfaces} 
Consider a Cauchy data set $(M,g_{ij},$ $K_{ij})$ as above and let $N$ be a
compact surface in
$M$ with normal $\eta$, second fundamental form $A$ and mean curvature
$H$. Then considering $N$ as a surface in an ambient Lorentzian space $V$
containing $M$, $N$ has two null normal fields which after a rescaling can be
taken to be $L_{\pm} = T \pm \norm$. 
Here $T$ is the future directed timelike unit normal of $M$ in $V$. 
The
null mean curvatures (or null expansions) corresponding to $L_{\pm}$ 
can be defined in terms of the variation of the area element $\mu_N$ of $N$ as 
$\delta_{L_{\pm}} \mu_N = \theta_{\pm} \mu_N$ or 
$$
\theta_\pm = \tr_N K \pm H,
$$
where $\tr_N K$ denotes the trace of the projection of $K_{ij}$ to $N$. 
Suppose $L_+$ is the outgoing null normal. 
$N$ is
called outer trapped (marginally trapped, untrapped) 
if $\theta_+ < 0$  ($\theta_+ = 0$, $\theta_+ > 0$). An
asymptotically flat spacetime which contains a trapped surface with $\theta_-
< 0$, $\theta_+ < 0$ is causally
incomplete. In the following we will for simplicity drop the word outer from
our terminology. 

Consider a Cauchy surface $M$. The boundary of the region in $M$
containing trapped surfaces is, if it is sufficiently smooth, a marginally
trapped surface. 
The equation $\theta_+ = 0$ is an equation analogous to the prescribed
curvature equation, in particular it is a 
quasilinear elliptic equation of second order.  
Marginally trapped surfaces 
are not variational in the same sense as minimal
surfaces. 
Nevertheless, they are stationary with respect to variations of
area within the outgoing light cone. 
The second variation of area along the outgoing null cone is given, 
in view of the Raychaudhuri equation, by 
\begin{equation}\label{eq:W}
\delta_{\phi L_+} \theta_+ = - ( G_{++} + |\sigma_+|^2 ) \phi ,
\end{equation}
for a function $\phi$ on $N$. Here $G_{++} = G_{\alpha\beta} L_+^\alpha
L_+^\beta$, and $\sigma_{+}$
denotes the shear of $N$ with respect to $L_+$, i.e. the trace-free part of
the null second fundamental form with respect to $L_+$. 
Equation (\ref{eq:W}) shows that the stability operator in the direction 
$L_+$ is not elliptic. 

In case of time-symmetric data, $K_{ij} = 0$, 
the dominant energy condition implies $\Scal_M \geq 0$
and marginally trapped surfaces are simply minimal surfaces. 
A stable compact 
minimal 2-surface $N$ in a 3-manifold $M$ 
with nonnegative scalar curvature 
must satisfy 
$$
2\pi \chi(N) = \int \kappa \geq \half \int_N \Scal_M + |A|^2 \geq 0
$$
and hence by the Gauss-Bonnet theorem, $N$ is diffeomorphic to a sphere
or a torus. In case $N$ is a stable minimal torus, 
the induced geometry is flat and the
ambient curvature vanishes at $N$.
If, in addition, $N$ minimizes, then $M$ is flat 
\cite{cai:galloway:torus}. 

For a compact marginally trapped surface  $N$ in $M$, 
analogous results can be
proved by studying the stability
operator defined with respect to the direction $\norm$. Let $J$ be the
operator defined in terms of a variation of $\theta_+$ by 
$J\phi = \delta_{\phi \norm}\theta_+$. Then 
\begin{align*}
J\phi = - \Delta_N \phi 
+ 2 s^A D_A \phi +  \left (\frac{1}{2} \Scal_N - s_A s^A 
 + D_A s^A   -\half |\sigma_+|^2 - G_{+-} \right ) \phi.
\end{align*} 
Here $s_A = - \frac{1}{2} \la L_-, D_A L_+ \ra$ and $G_{+-}$ is the Einstein
tensor evaluated on $L_+, L_-$. We may call $N$ stable if
the real part of the spectrum of $J$ is nonnegative. A sufficient condition
for $N$ to be stable is that $N$ is locally outermost. This can be formulated
for example by requiring that a neighborhood of $N$ in $M$
contains no trapped surfaces exterior to
$N$. In this case, assuming
that the dominant energy condition holds, $N$ is a sphere or a torus, and if
the real part of the spectrum of $J$ is positive then $N$ is a sphere. 
If $N$ is a torus, then the ambient curvature and shear vanishes at $N$, 
$s_A$ is a gradient, and $N$ is flat. One expects that in
addition, global rigidity should hold, 
in analogy with the minimal surface case. 
This is an open problem. If $N$ satisfies the stronger condition of strict
stability, which corresponds to the spectrum of $J$ having positive real
part, then $N$ is in the interior of a hypersurface $H$ of the ambient
spacetime, with the property that it is foliated by marginally trapped
surfaces \cite{AMS}. 
If the NEC holds and $N$ has nonvanishing shear, then $H$ is
spacelike at $N$. A hypersurface $H$ with these properties is known as a
dynamical horizon. 
\mnote{followed up by topological censorship etc} 

\subsection{Jang's equation} \label{sec:Jang}
Consider a Cauchy data set $(M,g_{ij},K_{ij})$. 
Extend $K_{ij}$ to a tensor field on $M
\times \Re$, constant in the vertical direction. Then the equation for a
graph 
$$
N = \{ (x,t) \in M \times \Re, \quad t = f(x) \}
$$
such that $N$ has mean curvature equal to the trace of the projection of
$K_{ij}$ 
to $N$ with respect to the induced
metric on $N$, is given by 
\begin{equation}\label{eq:jangs}
\sum_{i,j} 
\left ( K^{ij} - \frac{\nabla^i \nabla^j f}{(1+|\nabla
f|^2)^{1/2}} \right ) 
\left ( g_{ij} - \frac{\nabla_i f \nabla_j f}{1 +
|\nabla f|^2} \right ) 
= 0 , 
\end{equation} 
an equation closely related to the equation $\theta_+ = 0$. 
Equation (\ref{eq:jangs}) was introduced by P.~S. Jang \cite{jang} as part of
an attempt to generalize the inverse mean curvature flow method of Geroch
from time-symmetric to general Cauchy data. 

Existence and regularity for Jang's equation were proved by Schoen and Yau
\cite{SYII} and used to generalize their proof of the positive mass theorem
from the case of maximal slices to the general case. The solution to Jang's
equation is constructed as the limit of the solution to a sequence of
regularized problems. The limit consists of a collection $N$ of submanifolds of
$M\times \Re$. In particular, component near infinity is a graph and has the
same mass as $M$. $N$ may contain vertical components which project onto
marginally trapped surfaces in $M$, and in fact these constitute the only
possibilities for blow-up of the sequence of graphs used to construct $N$. 
If the DEC is valid, the metric on $N$ has non-negative scalar curvature in
the weak sense that 
$$
\int_N \Scal_N \phi^2 + 2|\nabla\phi|^2 \geq 0
$$
for smooth compactly supported functions $\phi$. If the DEC holds strictly,
the strict inequality holds and in this case the metric on $N$ 
is conformal to a metric with vanishing scalar curvature.

Jang's equation can be applied to prove existence of marginally trapped
surfaces, given barriers. 
Let $(M,g_{ij},K_{ij})$ be a
Cauchy data set 
containing two compact surfaces $N_1,
N_2$ which together bound a compact region $M'$ in $M$. 
Suppose the surfaces $N_1$ and
$N_2$ have $\theta_+ < 0$ on $N_1$ and $\theta_+ > 0$ on $N_2$. 
Schoen recently proved the following result.
\mnote{add some reference for Schoen's result}  
\begin{thm}[Existence of marginally trapped surfaces] 
\label{thm:schoen} 
Let $M',N_1, N_2$ be as above. Then there is a finite collection of compact,
marginally trapped surfaces
$\{\Sigma_a\}$ contained in the interior of $M'$, 
such that $\cup \Sigma_a$ is homologous to $N_1$.
If the DEC holds, then $\{\Sigma_a\}$ is a
collection of spheres and tori.
\end{thm} 
The proof proceeds by solving a
sequence of Dirichlet boundary value problems for Jang's equation with
boundary value on $N_1, N_2$ tending to $-\infty$ and $\infty$
respectively. The assumption on $\theta_+$ is used to show the existence of
barriers for Jang's equation.
Let $f_k$ be the sequence of solutions to the Dirichlet problems. 
Jang's equation is invariant under renormalization $f_k \to f_k + c_k$ 
for some sequence $c_k$ of real numbers. A 
Harnack inequality for the gradient of the solutions to Jang's equation is
used to show that the sequence of solutions $f_k$, 
possibly after a renormalization has a subsequence converging to a vertical
submanifold of $M' \times \Re$, which projects to a collection $\Sigma_a$ of
marginally trapped surfaces. By construction, the zero sets of the $f_k$ are
homologous to $N_1$ and $N_2$. The estimates on the sequence $\{f_k\}$ show
that this holds also in the limit $k \to \infty$. 
\mnote{in the appropriate homology class} 
The statement about the topology of the $\Sigma_a$ follows by showing,
using the above mentioned inequality for $\Scal_N$, that if DEC holds, 
the total Gauss curvature of each surface $\Sigma_a$ is non-negative.

\subsection{Center of mass} \label{sec:round} 
Since by the positive mass theorem $m > 0$ unless the ambient spacetime is
flat, it makes sense to consider the problem of finding an appropriate notion
of center of mass. This problem was solved by Huisken and Yau
who showed that under the asymptotic conditions
(\ref{eq:asymptcond}) the isoperimetric problem has a unique solution if one
considers sufficiently large spheres. 

\begin{thm}[Huisken and Yau \cite{huisken:yau}] \label{thm:HY} 
There is a $H_0 > 0$ and a compact region $B_{H_0}$ such that for each $H \in
(0,H_0)$ there is a unique constant mean curvature sphere $S_H$ with mean curvature
$H$ containted in $M \setminus B_{H_0}$. The spheres  form a foliation. 
\end{thm} 
The proof involves a study of the
evolution equation 
\begin{equation}\label{eq:flow} 
\frac{dx}{ds} = (H - \bar H) \norm
\end{equation} 
where $\bar H$ is the
average mean curvature. This is the gradient flow for the isoperimetric 
problem of
minimizing area keeping the enclosed volume constant. 
The solutions in
Euclidean space are standard spheres. Equation (\ref{eq:flow}) defines a
parabolic system, in particular we have 
$$
\frac{d}{ds} H = \Delta H + (\Ric(\norm,\norm) + |A|^2) (H - \bar H) .
$$
It follows from the fall-off conditions (\ref{eq:asymptcond}) that the
foliation of spheres constructed in Theorem \ref{thm:HY} are untrapped
surfaces. They can therefore be used as outer barriers in the existence
result for marginally trapped surfaces, Theorem  
\ref{thm:schoen}.

The mean curvature flow for a spatial hypersurface in a Lorentz manifold is
also parabolic. This flow has been applied  to construct constant mean
curvature Cauchy hypersurfaces in spacetimes.

\section{Maximal and related surfaces}
Let $N$ be the hypersurface $x_0 = u(x_1, \dots, x_n)$ in Minkowski space
$\Re^{1+n}$ with line element $-dx_0^2 + dx_1^2 + \cdots + dx_n^2$. Assume
$|\nabla u| < 1$ so that $N$ is spacelike. Then $N$ is
stationary with respect to variations of area if $u$ solves the equation 
\begin{equation}\label{eq:Lgraph}
\sum_i \nabla_i \left (\frac{\nabla_i u}{\sqrt{1 - |\nabla u|^2}} \right ) = 0
\end{equation} 
$N$ maximizes area with respect to compactly supported variations, and
hence is called a {\em maximal surface}. As in the case of the minimal
surface equation, equation (\ref{eq:Lgraph}) and more generally the
Lorentzian prescribed mean curvature equation, is quasilinear elliptic, but
it is not uniformly elliptic, which makes the regularity theory more subtle. 

A Bernstein principle analogous to the one for the minimal surface equation 
holds for the maximal surface equation
(\ref{eq:Lgraph}). 
Suppose that $u$ is a solution to (\ref{eq:Lgraph}) which is defined on all
of $\Re^n$. Then $u$ is an affine function
\cite{cheng:yau:lorentz}. 
An important tool used in the proof
is a Bochner type identity, originally due to Calabi, 
for the norm of the second
fundamental form. For a hypersurface in a flat ambient space, the Codazzi
equation states
$\nabla_i A_{jk} - \nabla_j A_{ik} = 0$. This gives the identity 
\begin{equation}\label{eq:calabi}
\Delta A_{ij} = \nabla_i\nabla_j H + A_{km} R_{\ i \ j}^{m\ k} + A_{mi}
\Ric_{\ j}^m
\end{equation} 
The curvature terms can be rewritten in terms of $A_{ij}$ if the ambient
space is flat. 
Using (\ref{eq:calabi}) to compute $\Delta |A|^2$ gives an expression which
is 
quadratic in $\nabla A$, and fourth order in $|A|$, and which 
allows one to perform
maximum principle estimates on $|A|$. 
Generalizations of this technique for hypersurfaces in general ambient
spaces play an important role in the proof of regularity of minimal surfaces,
and in the proof of existence
for Jang's equation, see section \ref{sec:Jang}, 
as well as in the analysis of the mean curvature flow
used to prove existence of round spheres, see section \ref{sec:round}. 
The generalization of equation (\ref{eq:calabi}) 
is known as a Simons identity.

For the case of maximal hypersurfaces of Minkowski space, it follows from 
further maximum principle estimates, that a maximal hypersurface of
Minkowski space is convex, in particular it 
has non-positive Ricci curvature. Generalizations of this technique allows
one to analyze entire constant mean curvature hypersurfaces of Minkowski
space.

Consider a globally hyperbolic
Lorentzian manifold $(V,\gamma)$.
A $C^0$ hypersurface 
is said to be weakly spacelike if timelike
curves intersect it in at most one point. 
Call a codimension
two submanifold $\Gamma
\subset V$ a weakly spacelike boundary if it bounds a weakly spacelike
hypersurface $N_0$. 
\begin{thm}[Existence for Plateau problem for maximal surfaces 
\cite{bartnik:acta}] 
Let $V$ be a globally hyperbolic spacetime and assume the causal structure of
$V$ is such that the domain of
dependence of any compact domain in $V$ is compact. 
Given a weakly spacelike boundary $\Gamma$ in $V$, there is a weakly spacelike
maximal hypersurface $N$ with $\Gamma$ as its boundary. $N$ is
smooth except possibly on null geodesics connecting points of $\Gamma$.
\end{thm} 
Here maximal hypersurface is understood in a weak sense, referring to
stationarity with respect to variations. Due to the non-uniform ellipticity
for the maximal surface equation, the interior regularity which holds for
minimal surfaces fails to hold in general for the maximal surface equation. 

A time oriented spacetime is said to have a crushing singularity to the
past (future) 
if there is a sequence
$\Sigma_n$ of Cauchy surfaces so that the mean curvature function $H_n$ of 
$\Sigma_n$, diverges uniformly to $-\infty$ ($\infty$). 

\begin{thm}[Gerhardt \cite{gerhardt:H-surfaces}] \label{thm:gerhardt} 
Suppose that $(V,\gamma)$ is globally hyperbolic with compact Cauchy surfaces 
and satisfies the SEC. Then if
$(V,\gamma)$ 
has crushing singularities to the past and future it
is globally foliated
by constant mean curvature hypersurfaces. The mean curvature 
$\tau$ of these Cauchy surfaces is a global time
function. 
\end{thm} 
The proof involves an application of results from geometric measure theory 
to an action $\Energy$ of the form discussed in section \ref{sec:minimal}. 
A barrier argument is used to control the maximizers.
Bartnik \cite[Theorem 4.1]{bartnik:AF} gave a direct proof of existence of a
constant mean curvature (CMC) hypersurface, given barriers. 
If the spacetime $(V,\gamma)$ is symmetric, so that a compact 
Lie group acts on $V$ by
isometries, 
then CMC hypersurfaces in $V$ inherit the symmetry.
Theorem \ref{thm:gerhardt} 
gives a condition under which a spacetime is globally foliated by
CMC hypersurfaces. 
In
general, if the SEC holds in a spatially compact spacetime, then for each
$\tau \ne 0$, there is at most one constant mean curvature Cauchy surface
with mean curvature $\tau$. 
In case $V$ is vacuum,
$\Ric_V = 0$, and 3+1 dimensional, then each point $x \in V$ is on at most
one hypersurface of constant mean curvature unless $V$ is flat and splits as
a metric product. 
\mnote{add cite for this [brill:flaherty:76,marsden:tipler]} 

There are vacuum spacetimes with compact Cauchy surface which contain no CMC
hypersurface \cite{chrusciel:etal:gluing}. The proof is carried out by
constructing Cauchy data, using a gluing argument, 
on the connected sum of two tori, 
such that the resulting Cauchy data set $(M,g_{ij}, K_{ij})$ 
has an involution which reverses the
sign of $K_{ij}$. The involution extends to the maximal vacuum
development $V$ of the Cauchy data set. Existence of a CMC surface in $V$
gives, in view of the involution, barriers which allow one to construct a
maximal Cauchy surface homeomorphic to $M$. 
This leads to a contradiction, since  the connected sum of two
tori does not carry a metric of positive scalar curvature, and therefore, in
view of the constraint equations, cannot be imbedded as a maximal Cauchy
surface in a vacuum spacetime. 
\mnote{in view of the Gannon type singularity theorem
[galloway:gannon]} 
The maximal vacuum development $V$ is causally
geodesically incomplete. However, in view of the existence
proof for CMC Cauchy surfaces, cf. Theorem \ref{thm:gerhardt}, 
these spacetimes cannot have a
crushing singularity. 
It would be 
interesting to settle the open question whether there are stable 
examples of this type. 

In the case of a spacetime $V$ which has an expanding end,
one does not expect in general that the spacetime is globally foliated by CMC
hypersurfaces even if $V$ is vacuum and contains a CMC Cauchy surface. 
This expectation is based on the
phenomenon 
known as the collapse of the lapse; for example the Schwarzschild spacetime
does not contain a global foliation by maximal Cauchy surfaces
\cite{beig:omurchadha:98}. However, no counterexample is known in the
spatially compact case.  
In spite of these caveats, many
examples of spacetimes with global CMC foliations are known and the CMC
condition, or more generally prescribed mean curvature 
is an important gauge condition for general relativity. 

Some examples of situations where global constant or prescribed mean
curvature foliations are known to
exist in vacuum or with some types of matter 
are spatially homogenous spacetimes spacetimes, and spacetimes with two
commuting Killing fields. 
Small
data global existence for the Einstein equations with CMC time gauge have
been proved for  
spacetimes with one Killing field, with Cauchy
surface a circle bundle over a surface of genus $> 1$, by Choquet-Bruhat and
Moncrief.  
Further, for 
3+1 dimensional spacetimes with Cauchy surface admitting a hyperbolic metric,
small data global existence in the expanding direction has been proved by
Andersson and Moncrief. 
See 
\cite{andersson:global} and \cite{rendall:survey} for surveys on 
the Cauchy problem in general relativity. 

\mnote{say something about maximal and CMC hypersurfaces in asymptotically
flat spacetimes} 

\section{Null hypersurfaces} 
Consider an asymptotically flat spacetime containing a black hole, i.e. a
region $B$ such that future causal curves starting in $B$ cannot reach
observers at infinity. 
The boundary of the trapped region is called the event horizon
$\Hor$. This is a null hypersurface, which under reasonable conditions on
causality has 
null generators which are complete to the future. Due to the completeness,
assuming that $\Hor$ is smooth, 
one can use the Raychaudhuri equation (\ref{eq:W}) to show that the null
expansion $\theta_+$ of a spatial cross section of $\Hor$ must satisfy
$\theta_+ \geq 0$, and hence that the area of cross sections of $\Hor$ grows
monotonously to the future. A related
statement is that null generators can enter $\Hor$ but may not leave it.  
This was first proved by 
Hawking for the case of smooth horizons, using
essentially the Raychaudhuri equation. 
In general $\Hor$ can fail to be smooth. However, from the definition of
$\Hor$ as the boundary of the trapped region follows that it has support
hypersurfaces, which are past lightcones. 
This property allows one to prove
that $\Hor$ is Lipschitz and hence smooth 
almost everywhere. At smooth
points of $\Hor$, the calculations in the proof of Hawking applies, and the
monotonicity of the area of cross sections follows. 
\mnote{see also later paper by Chrusciel, Fu and others} 

\begin{thm}[Area theorem \protect{\cite{chrusciel:etal:areathm}}]
Let $\Hor$ be a black hole event horizon in a smooth spacetime
$(M,g)$. Suppose that the generators are future complete and the N.E.C. holds
on $\Hor$. Let $S_a$, $a=1,2$ be two spacelike cross sections of $\Hor$ and
suppose that $S_2$ is to the future of $S_1$. Then $\Area(S_2) \geq
\Area(S_1)$. 
\end{thm} 

The eikonal equation $\nabla^\alpha u \nabla_\alpha u = 0$ plays a central
role in geometric optics. Level sets of a solution $u$ are null hypersurfaces
which correspond to wave
fronts. Much of the recent progress on rough solutions to 
the Cauchy problem for quasilinear
wave equations
is based on understanding 
the influence of the geometry of these wave fronts
on the evolution of high-frequency modes in the background spacetime. 
In this analysis many objects familiar from general relativity, such as 
the structure equations for null hypersurfaces, the Raychaudhuri equation, 
and the Bianchi identities play an important role, together with novel
techniques of geometric analysis used to control the geometry of cross
sections of the wave fronts and to estimate the connection coefficients in a
rough spacetime geometry. These techniques show great promise 
and can be expected 
to have a significant impact on our understanding of the Einstein equations
and general relativity.

\subsection*{Acknowledgements} The author is grateful to, among others, Greg
Galloway, 
Gerhard Huisken, Jim Isenberg, Piotr Chrusciel and Dan Pollack
for helpful discussions concerning the topics covered in this article.

\providecommand{\bysame}{\leavevmode\hbox to3em{\hrulefill}\thinspace}
\providecommand{\MR}{\relax\ifhmode\unskip\space\fi MR }
\providecommand{\MRhref}[2]{%
  \href{http://www.ams.org/mathscinet-getitem?mr=#1}{#2}
}
\providecommand{\href}[2]{#2}


\end{document}